\documentclass[pdflatex,sn-mathphys-num]{sn-jnl}

\usepackage{graphicx}%
\usepackage{multirow}%
\usepackage{amsmath,amssymb,amsfonts}%
\usepackage{amsthm}%
\usepackage{mathrsfs}%
\usepackage[title]{appendix}%
\usepackage{xcolor}%
\usepackage{textcomp}%
\usepackage{manyfoot}%
\usepackage{booktabs}%
\usepackage{algorithm}%
\usepackage{algorithmicx}%
\usepackage{algpseudocode}%
\usepackage{listings}%
\usepackage{mathrsfs}%

\theoremstyle{thmstyleone}%
\theoremstyle{thmstyletwo}%

\theoremstyle{thmstylethree}%

\begin{document}

\title[Article Title]{Bringing Computation to the data: Interoperable serverless function execution for astrophysical data analysis in the SRCNet}


\author*[1]{\fnm{Manuel} \sur{Parra-Royón}}\email{mparra@iaa.es} 
\author[1]{\fnm{Julián} \sur{Garrido-Sánchez}}\email{jgarrido@iaa.csic.es} 
\author[1]{\fnm{Susana} \sur{Sánchez-Expósito}}\email{sse@iaa.es}
\author[1]{\fnm{María Ángeles} \sur{Mendoza}}\email{amendoza@iaa.es}
\author[2]{\fnm{Rob} \sur{Barnsley}}\email{rbarnsley@skao.int}
\author[4]{\fnm{Anthony} \sur{Moraghan}}\email{anthony.moraghan@manchester.ac.uk}
\author[1]{\fnm{Jesús} \sur{Sánchez}}\email{jsanchez@iaa.es}
\author[1]{\fnm{Laura} \sur{Darriba}}\email{ldarriba@iaa.es}
\author[3]{\fnm{Carlos} \sur{Ruíz-Monje}}\email{riosmonjecarlos@gmail.com}
\author[1]{\fnm{Edgar} \sur{Joao}}\email{edgar@iaa.es}
\author[1]{\fnm{Javier} \sur{Moldon}}\email{jmoldon@iaa.es}
\author[2]{\fnm{Jesús} \sur{Salgado}}\email{j.salgado@skao.int}
\author[1]{\fnm{Lourdes} \sur{Verdes-Montenegro}}\email{lourdes@iaa.es}

\affil*[1]{\orgdiv{Extragalatic Astronomy}, \orgname{Instituto de Astrofísica de Andalucía}, \orgaddress{\street{Glorieta de la Astronomía, s/n}, \city{Granada}, \postcode{18008}, \state{Andalucía}, \country{Spain}}}
\affil*[2]{\orgdiv{Square Kilometre Array Observatory}, \orgname{SKAO}, \orgaddress{\street{Jodrell Bank, Lower Withington}, \city{ Macclesfield}, \postcode{SK11 9FT}, \state{Cheshire}, \country{United Kingdom}}}
\affil*[3]{\orgdiv{Theoretical Physics, Multidisciplinary Unit for Energy Science}, \orgname{University of Sevilla}, \orgaddress{\city{Sevilla}, \postcode{41080}, \state{Andalucía}, \country{Spain}}}
\affil*[4]{\orgdiv{Jodrell Bank Centre for Astrophysics. School of Physics and Astronomy}, \orgname{University of Manchester}, \orgaddress{\city{Manchester}, \postcode{M13 9PL}, \country{United Kingdom}}}

\abstract{
\textbf{Purpose:} 
Serverless computing is a paradigm in which the underlying infrastructure is fully managed by the provider, enabling applications and services to be executed with elastic resource provisioning and minimal operational overhead. A core model within this paradigm is Function-as-a-Service (FaaS), where lightweight functions are deployed and triggered on demand, scaling seamlessly with workload. FaaS offers flexibility, cost-effectiveness, and fine-grained scalability, qualities particularly relevant for large-scale scientific infrastructures where data volumes are too large to centralise and computation must increasingly occur close to the data. The Square Kilometre Array Observatory (SKAO) exemplifies this challenge. Once operational, it will generate about 700~PB of data products annually, distributed across the SKA Regional Centre Network (SRCNet), a federation of international centres providing storage, computing, and analysis services. In such a context, FaaS offers a natural mechanism to bring computation to the data.  

\textbf{Methods:} 
We studied the principles of serverless and FaaS computing and explored their application to radio astronomy workflows. Representative functions for astrophysical data analysis were developed and deployed, including micro-functions derived from existing libraries and wrappers around domain-specific applications. In particular, a \emph{Gaussian convolution} function was implemented and integrated within the SRCNet ecosystem to validate feasibility.  

\textbf{Results:} 
The use case demonstrates that FaaS can be embedded into the existing SRCNet architecture and ecosystem of services, allowing functions to run directly at sites where data replicas are stored. This reduces latency, minimises long-distance transfers, and improves efficiency, aligning with the SRCNet vision of federated, data-proximate computation.  

\textbf{Conclusions:} 
The results show that serverless models, when integrated with federated infrastructures such as the SRCNet, provide a scalable and efficient pathway to address the data volumes of the SKA era.  

}


\maketitle

\section{Introduction}
\label{sec:introduction}

Over the past two decades, scientific computing has undergone successive paradigm shifts to meet the demands of increasingly data-intensive research \cite{kim2009cloud}. Early client–server models gave way to grid computing, High-Performance computing (HPC), and, more recently, large-scale cloud infrastructures that abstract physical resources into elastic, on-demand services \cite{armbrust2010view,marinescu2017cloud,fox2009above}. Within this landscape, \emph{serverless computing} has emerged as a model in which the infrastructure is fully managed by the provider, resources are provisioned automatically, and execution is billed at a fine-grained level \cite{castro2019rise,hellerstein2018serverless}. 

\begin{figure}[h]
\centering
\includegraphics[width=1\textwidth]{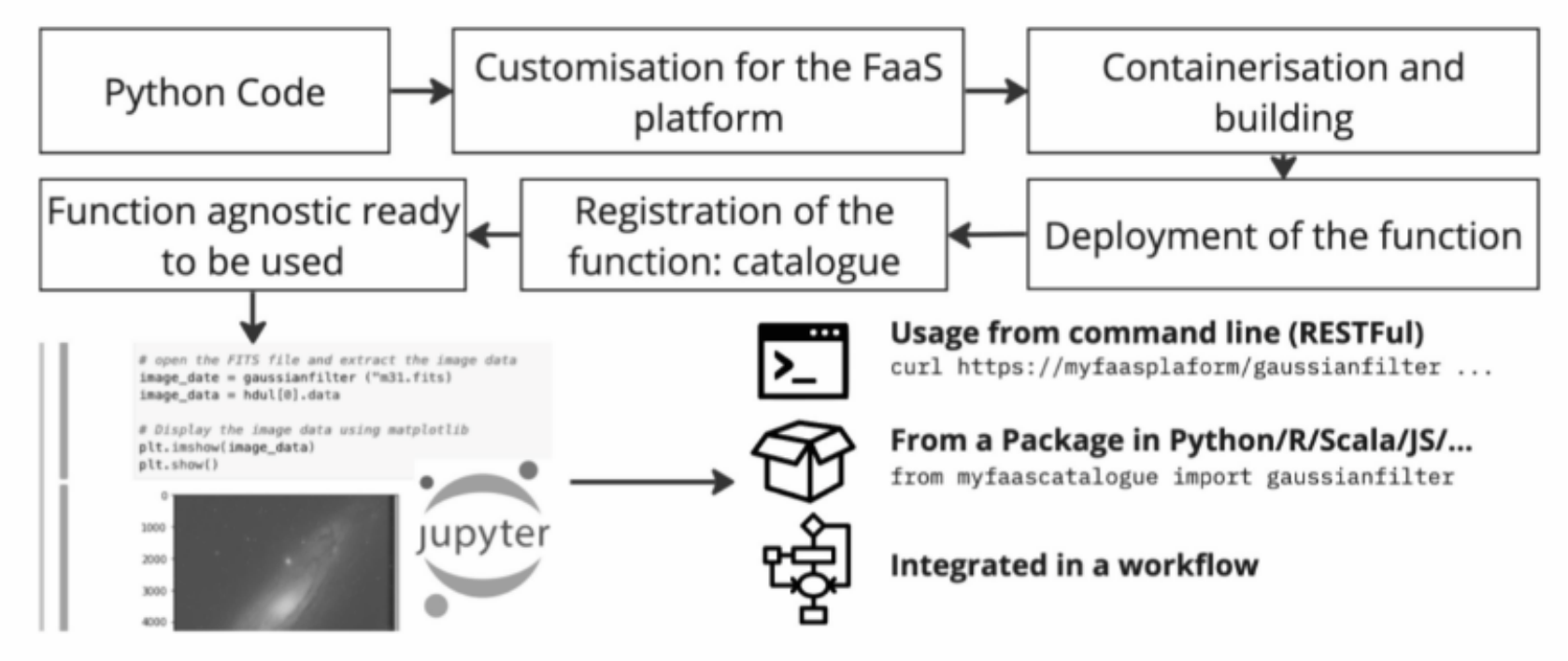}
\caption{Model of a pipeline to enable a function on a FaaS environment, that encompasses everything from the creation of the Python code for the function to its direct use in a Notebook, including building, deployment, and registration in a catalogue.}\label{fig:pipeline-serverless}
\end{figure}

A central paradigm of serverless is \emph{Function-as-a-Service} (FaaS), in which developers write lightweight, stateless functions that are triggered on demand by events or requests. These functions execute in isolated containers for the duration of the task, after which the resources are released. This model removes the burden of server provisioning and scaling, enabling developers and scientists to focus entirely on logic and application design. Commercial platforms such as Amazon Web Servces (AWS) Lambda, Google Cloud Functions, and Azure Functions exemplify this model in industry, while open-source alternatives such as OpenFaaS, Knative, and Apache OpenWhisk bring the paradigm into private or hybrid infrastructures \cite{djemame2020openwhisk,le2022openfaas,goodwin2023knative}. Figure \ref{fig:pipeline-serverless} describes the process for building a generic function in FaaS. In this context, FaaS offers cost-effectiveness, resilience, and elasticity, and has already demonstrated its ability to drive parallelism in scientific workloads through frameworks such as \texttt{PyWren} \cite{pywren2017}, \texttt{numpywren} \cite{numpywren2018}, and \texttt{Lithops}  \cite{lithops_docs,arjona2023transparent}. Yet its systematic integration into large-scale scientific infrastructures remains relatively under-explored \cite{malawski2023serverless}. 

The promise of FaaS is particularly relevant in the context of data-intensive sciences, where datasets are too large to move freely across networks. Traditional workflows that transfer raw data to central processing sites are increasingly infeasible at petascale volumes. Instead, \emph{data-proximate computation}---deploying code close to where the data reside---has become a guiding principle for modern infrastructures \cite{anderson2010data,rosenthal2014cloud}. FaaS naturally aligns with this model: functions are self-contained and portable, and can be deployed at the sites where data are stored, reducing both network traffic and latency while improving efficiency and throughput across distributed infrastructures.


The Square Kilometre Array Observatory (SKAO) exemplifies these challenges. As one of the most ambitious radio-astronomy projects of the 21st century, it will generate on the order of 700~PB of data products annually, with unprecedented sensitivity and resolution \cite{hartley2023ska}. Exploiting this data requires not only high-performance pipelines for calibration, imaging, and spectral analysis, but also a federated infrastructure for global access. This role is fulfilled by the SKA Regional Centre Network (SRCNet), a distributed federation of regional centres that together provide storage, computation, staging, and analysis services for the international community \cite{broekema2020ska,salgadoska}. In this environment, each SRC node operates as both a repository of data replicas provided by the SKAO Rucio Datalake, and a computing hub, and collectively the SRCNet must deliver scalable, secure, and interoperable access to resources across heterogeneous environments. This makes it a natural testbed for leading-edge paradigms such as FaaS, where the concept of \textit{data-proximate computation}, or “moving software to the data”, is increasingly recognised as a cornerstone for next-generation scientific infrastructures \cite{anderson2010data,rosenthal2014cloud}. 

Serverless computing can be highly beneficial for SRCNet data processing capabilities due to its scalability, cost-effectiveness, efficiency, and reliability. With serverless models, users can elastically discover and scale resources ---defined as functions or micro-functions--- as needed, ensuring that SKA workflows remain highly available. Functions exposed can be used through APIs---i.e. by using OpenAPI\footnote{OpenAPI: \url{https://spec.openapis.org/oas/latest.html}}---and integrated seamlessly into diverse environments, from Jupyter Notebooks to workflow engines and command-line tools, enabling reproducible and portable distributed pipelines. In radio-astronomy and in the context of SKA, these functions may encapsulate operations from precursors such as MeerKAT \cite{booth2012overview}, ASKAP \cite{johnston2008science}, or MWA \cite{lonsdale2009murchison}, spanning tasks like image filtering, cube visualisation, spectral analysis, and source extraction, among others (see Figure \ref{fig:serverless-gateway}, which shows how different operations and versions of these operations can be used). Serverless thereby provides not only an abstraction for executing code, but a unifying layer through which scientific communities can interact, share, reuse, and orchestrate distributed computing at scale.

This work advances this line of research by implementing scientific operations through a function-oriented architecture that can be natively integrated into the existing and operational SRCNet service ecosystem. First, we analyzed different platforms for the development of FaaS services and applied this model to a scientific workflow that includes the most common steps in radio-astronomy data analysis. We then validated the function-oriented architecture within the SRCNet service ecosystem by implementing a \emph{Gaussian convolution} as a representative scientific operation. These operations are containerized with APIs and deployed on Kubernetes clusters running on SRCNet nodes. Finally, SRCNet global services are configured to provide authenticated access to these operations. 


The structure of this article is as follows. Section~\ref{sec:relatedworks} reviews related research on the use of serverless and FaaS platforms. Section~\ref{sec:analysis-faas} provides an overview of existing FaaS platforms, while Section~\ref{sec:faas-astro} introduces their application to astrophysical workflows. Section~\ref{sec:faas-srcnet-context} describes the SRCNet service ecosystem and presents  the implementation of a FaaS use case within this framework. The implications of this model are discussed in Section~\ref{sec:discussion}, and Section~\ref{sec:conclusion} summarises our conclusions and outlines directions for future work.

\section{Related works}
\label{sec:relatedworks}

The evolution of distributed computing models has progressively reduced the granularity of the execution unit, from virtual machines (VM) to containers and, more recently, functions. Virtualisation technologies and IaaS stacks such as OpenStack and OpenNebula \cite{ibm_hybrid_2024,openstack_hpc_booklet} enabled research centres to provision VM-based resources with capabilities comparable to commercial clouds while retaining policy control and integration with scientific storage and networks \cite{openstack_analysis_2015,razavi2015opennebula,huang2013evaluation}. This VM-centric model provided elasticity through features such as CPU and RAM overcommitment\footnote{OpenStack architecture and the overcommit design of the scheduler: \url{https://docs.openstack.org/arch-design/design-compute/design-compute-overcommit.html}}, but remained relatively heavy for fine-grained or irregular workloads.

Containers and orchestrators such as Kubernetes introduced a lighter abstraction, with rapid deployment, scaling, and support for microservices \cite{jonas2019berkeley}. Building on this substrate, serverless computing emerged as a model in which the infrastructure is fully managed by the provider, resources are provisioned automatically, and execution is billed at fine granularity \cite{castro2019rise,hellerstein2018serverless}. In FaaS, the dominant form of serverless, developers deploy functions that are triggered by events or requests, executed in isolated containers, and terminated upon completion. This model removes the burden of provisioning and scaling, allowing developers and scientists to focus on application logic. Major cloud providers (AWS Lambda, IBM Cloud Functions, Azure Functions) exemplify the model, while open-source alternatives (OpenFaaS \cite{mohanty2018serverless,le2022openfaas}, Knative \cite{knative_docs,goodwin2023knative}, Apache OpenWhisk \cite{djemame2020openwhisk,kuntsevich2018openwhisk}) extend it to hybrid and private infrastructures \cite{djemame2020openwhisk,le2022openfaas,goodwin2023knative}.

FaaS has demonstrated its ability to drive massive parallelism in data-intensive tasks. \texttt{PyWren} showed how unmodified Python code could be mapped to AWS Lambda for elastic, highly parallel workloads \cite{pywren2017}. Building on this, \texttt{numpywren} applied serverless to dense linear algebra, achieving performance within 33\% of ScaLAPACK for some kernels \cite{numpywren2018,numpywren2020}. IBM’s \texttt{Lithops} generalised these concepts into a multi-cloud toolkit for analytics atop serverless backends and object stores \cite{lithops_docs,arjona2023transparent}. At the federation level, \texttt{funcX} introduced a federated FaaS fabric that decouples a cloud-hosted control plane from site-hosted endpoints, enabling execution across clusters, HPC, and edge resources with routing strategies to minimise time-to-completion \cite{funcx_tpds_2022,funcx_arxiv_2022}.

Despite this progress, systematic integration of FaaS into large-scale science remains limited. Challenges include data access and locality, cold-start overheads \cite{goodwin2023knative,openfaas_autoscaling_2024}, state management, and security constraints \cite{mcgrath2017serverless,jonas2019berkeley}. Commercial ecosystems have also explored algorithm marketplaces (e.g., former Algorithmia, now DataRobot\footnote{DataRobot:  \url{https://www.datarobot.com/product/ai-platform/?redirect_source=algorithmia.com}}) where functions are published as APIs for reuse across communities \cite{arya2018api_marketplace}. In astronomy, interoperability has traditionally been addressed through the International Virtual Observatory Alliance (IVOA) standards, with protocols such as \emph{DataLink} that link datasets to access services and related processing operations \cite{ivoa_datalink_2015,ivoa_datalink_2023}. Leveraging such standards in combination with serverless fabrics offers a path to integrate computation and discovery.

The relevance of this model becomes evident when considering the scale of modern scientific instruments. Projects such as the Large Hadron Collider \cite{bird2011update} (LHC), LIGO \cite{abbott2016ligo}, and the SKA generate unprecedented data volumes that must be processed across international federated infrastructures \cite{bird2011update,abbott2016ligo,hartley2023ska}. Currently, scientific communities rely on distributed infrastructures that span multiple data centres, hybrid clouds, and HPC clusters to provide the computational power and storage capacity required. In this environment, both datasets and computational jobs are distributed across sites, and the cost of transferring data to a central location is prohibitive. 

An additional aspect discussed in recent studies concerns energy efficiency in Big Science. VM-centric approaches often lead to idle consumption due to coarse-grained resource allocation, whereas containers reduce this overhead but still require pre-provisioned services. FaaS, by allocating resources only during function execution, aligns energy use more closely with actual demand, offering a pathway to reduce the carbon footprint of large-scale infrastructures \cite{jonas2019berkeley,mcgrath2017serverless}.

In this context, FaaS provides a natural mechanism to bring computation to the data in an efficient way. By deploying functions close to where the data are generated or stored, FaaS reduces latency and network traffic, avoiding unnecessary transfers to central sites and enabling efficient exploitation of distributed resources. This principle---computation close to the data---is a cornerstone for future infrastructures such as the SRCNet.

\section{Analysis of FaaS platforms}
\label{sec:analysis-faas}

To evaluate the suitability of different FaaS frameworks, we considered a range of open-source and customisable platforms that can be deployed on Kubernetes clusters on a public Cloud Computing environment. Each of these platforms provides a different balance between ease of installation, language/runtime support, operational maturity, and integration with external services. Table~\ref{tab:faas-sideways} summarises the most relevant options gathered.

\begin{sidewaystable}
\caption{Comparison of FaaS/serverless platforms for the SRCNet testbed.}
\label{tab:faas-sideways}
\begin{tabular*}{\textheight}{@{\extracolsep\fill}lcccc}
\toprule
\textbf{Platform} & \textbf{Development status} & \textbf{Installation complexity} & \textbf{Kubernetes integration} & \textbf{Language support} \\
\midrule
OpenFaaS & Active, mature & Low & Native (Helm) & Go, Python, Node, C\#, Java \\
Knative & Active (CNCF) & Medium--High & Native (Ingress/Istio) & Polyglot (containers) \\
Kubeless & Maintenance mode & Low & Native & Python, Node, Ruby, Go \\
Fission & Active & Medium & Native & Python, Go, Node, Java, .NET \\
Apache OpenWhisk & Active (Apache) & High & Via Kafka/ZooKeeper & Java, Python, Node, Swift, etc. \\
Nuclio & Active & Medium & Native & Python, Go, Java, .NET, R \\
AWS Lambda & Active, commercial & N/A (managed) & Not K8s-native & Polyglot \\
Google Cloud Run & Active, commercial & N/A (managed) & Not K8s-native & Polyglot (containers) \\
Custom on Kubernetes & N/A (CI/CD) & Variable & Native (Deployments/Services) & Any (container) \\
\botrule
\end{tabular*}
\end{sidewaystable}

OpenFaaS remains one of the most widely adopted open-source solutions, with a large community and relatively simple Helm-based installation. Knative, as a CNCF\footnote{Cloud Native Computing Foundation: \url{https://www.cncf.io/}} project, offers stronger autoscaling and extensibility but requires additional ingress components. Kubeless, in contrast, is lightweight but now in maintenance mode, making it suitable for pilots but with limited long-term prospects. Fission provides a rich trigger model and moderate adoption, while Apache OpenWhisk offers polyglot support and scalability at the expense of a more complex installation involving Kafka and ZooKeeper. Nuclio is particularly suited to machine learning and GPU-accelerated workloads. AWS Lambda and Google Cloud Run are included as reference points due to their maturity and adoption in industry, but they are managed services and not directly deployable in SRCNet clusters. Finally, the ``Custom on Kubernetes'' model leverages Continuos Integration and Continuos Development (CI/CD) together with a Harbor registry, integrating directly with any existing  ecosystem to provide FaaS-like capabilities without relying on a dedicated serverless framework.

All of the evaluated platforms highlighted in Table \ref{tab:faas-sideways} require Kubernetes as the underlying cluster manager. This prerequisite ensures orchestration, autoscaling, monitoring, and high availability, and allows functions to be deployed as containers with consistent lifecycle management. 

\section{Application of FaaS model to a radio-astronomy workflow}
\label{sec:faas-astro}

Interferometric data analysis in radio-astronomy is typically organised as a chained workflow comprising data ingestion and manipulation, flagging, calibration, (optionally) splitting/averaging, imaging and deconvolution, and self-calibration, with instrument- or mode-specific variations. In the SKA/SRCNet context these workflows must be \emph{globally available} and callable from heterogeneous environments (Jupyter Notebooks, workflow engines, command-line tools, APIs or Python packages, among others), while operating close to distributed data replicas to minimise transfers. A serverless FaaS approach fits this requirement: it encapsulates atomic steps of the pipeline as simple, stateless functions that can be deployed where the data are stored, scaling elastically and exposing uniform APIs. This enables the community to containerise commonly used operations from precursor ecosystems (MeerKAT, ASKAP, MWA)---spanning image filtering, cube visualisation, spectral analysis, source extraction, etc.---and invoking them across regions as policy-compliant services \cite{booth2012overview,johnston2008science,lonsdale2009murchison}.

\begin{figure}[t]
  \centering
  \includegraphics[width=1\linewidth]{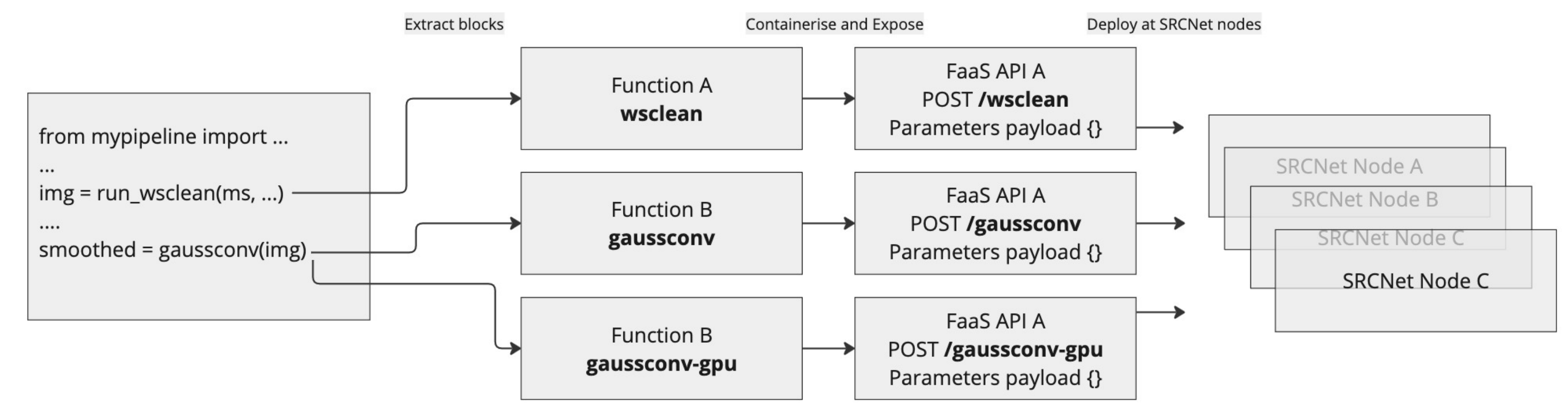}
  \caption{Decomposition of a radio-astronomy pipeline: a monolithic script with \texttt{wsclean} and \texttt{gaussconv} is decomposed into containerised functions and exposed as FaaS HTTP endpoints. Functions are then deployed at SRCNet nodes hosting the dataset replicas to enable data-proximate execution. The same function may have variants, for example with improvements in implementation such as GPU support \texttt{gaussconv-gpu}, which can even run on nodes enabled for this purpose.}
  \label{fig:pipeline-to-function}
\end{figure}

Many radio-astronomy tools already provide well-scoped operations suitable for function extraction. CASA tasks \cite{bean2022casa} (e.g., calibration and imaging primitives), standalone binaries such as \textit{wsclean} (imaging) and \textit{aoflagger} (flagging) \cite{offringa2014wsclean,offringa2010aoflagger,bean2022casa}, or purpose-built micro-functions can each be mapped to a single function with a simple HTTP interface (i.e. based on OpenAPI). Figure \ref{fig:pipeline-to-function} depicts the procedure for extracting from a monolith code to analyse data into an execution based on functions that can be shared across a distributed infrastructure. The extraction process consists of: (i) identifying inputs/outputs (e.g., MeasurementSets, FITS cubes, calibration tables, etc.); (ii) defining a parameter schema according to the operation; (iii) containerising the executable and dependencies; and (iv) declaring derived data products and outputs for downstream chaining.

Table~\ref{tab:faas-task-mapping} sketches representative pipeline steps and their function realisations. Section \ref{subsec:implementation} describes in more detail the Gauss Convolution (\texttt{gaussconv}) for clarity, but the same principles apply to CASA-based steps and domain binaries such as \textit{wsclean} and \textit{aoflagger}.

\begin{table}[h]
\centering
\caption{Examples of radio-astronomy tasks mapped to FaaS functions (indicative).}
\label{tab:faas-task-mapping}
\begin{tabular}{@{}p{0.22\linewidth}p{0.23\linewidth}p{0.22\linewidth}p{0.29\linewidth}@{}}
\toprule
\textbf{Pipeline step} & \textbf{Fn} & \textbf{I/O \& profile} & \textbf{Notes} \\
\midrule
Flagging & \texttt{aoflagger} & MS $\rightarrow$ MS  & Deterministic params; high I/O; site-local preferred. \\
Calibration & \texttt{casa-applycal} & MS + caltables $\rightarrow$ MS  & CASA task wrapper; provenance of tables required. \\
Imaging & \texttt{wsclean} & MS $\rightarrow$ image cube (GPU) & GPU-aware scheduling; large scratch I/O. \\
Smoothing & \texttt{gaussconv} & FITS $\rightarrow$ FITS  & Lightweight; good candidate for function chains. \\
Smoothing & \texttt{gaussconv-gpu} & FITS $\rightarrow$ FITS (GPU) & Lightweight; good candidate for function chains. \\
Source extraction & \texttt{pybdsf} & FITS $\rightarrow$ catalogue  & Produces VO tables; integrates with VO services. \\
\bottomrule
\end{tabular}
\end{table}
 
Overall, in a context where both datasets and jobs are geographically distributed, FaaS makes it possible to run functions exactly where the data reside, thereby minimising unnecessary transfers and improving end-to-end workflow efficiency.

\section{FaaS in the SRCNet context}
\label{sec:faas-srcnet-context}

The previous sections have outlined the potential of serverless computing and the mapping of radio-astronomy workflows into simple functions. We now turn to the practical integration of this paradigm within the SRCNet. Unlike commercial cloud providers, SRCNet operates as a distributed federation of regional centres geographically disseminated, where each SRC hosts data replicas and computational resources under local governance but coordinated through common standards and global services, this is, SRCNet local services and SRCNet Global services, respectively. This federated model imposes unique constraints on function execution: authentication and authorisation must be harmonised across institutions; data and functions must be registered and discoverable across sites; and execution must be located as close as possible to the relevant dataset replicas. 

The aim of this section is to show how function-oriented workflows can be realised in practice across a federated scientific infrastructure and leveraging the SRCNet services to provide secure, scalable, and data-proximate execution. Sub-section~\ref{subsec:srcnet-arch} presents the SRCNet architecture relevant to function deployment and orchestration, including the role of site-local and global services. Section~\ref{subsec:implementation} details a concrete implementation of the \emph{Gaussian convolution} function (\texttt{gaussconv}), illustrating how functions are containerised, registered, authorised, and exposed through the SRCNet local and global service stack. 

\subsection{The SRCNet Architecture}
\label{subsec:srcnet-arch}

The SRCNet is designed as a federated ecosystem in which both global and site-local services must interoperate seamlessly. Its architecture is being explicitly tailored to support large-scale scientific workloads, such as those required by SKA data processing. Within this framework, the deployment of serverless functions requires integration with both SRCNet global services, which define identity, authorisation, data distribution and service catalogues and discovery, and SRCNet local services, which provide the actual compute and storage resources at each SRC node.  

\paragraph{SRCNet Global Services}

At the federation level, several services orchestrate the consistent operation of the SRCNet:
\begin{itemize}
    \item SKAO Identity and Access Management (SKAO IAM), required for user authentication and role-based access control across the federation. It issues OIDC (OpenID Connect protocol) tokens that are consumed by services and functions to validate user identity and authorisation contexts.
    \item The SRCNet Permissions API builds on IAM to manage fine-grained authorisation policies, defining whether a user may access to specific datasets, functions, or services across one or more SRC nodes, enabling cross-site enforcement of data-governance rules.
    \item SRCNet Site-Capabilities serves as the federation-wide catalogue of services. Each SRC node publishes its available resources, functions, and capabilities through this registry, exposing metadata such as endpoints, ports, parameters, and associated storage.
    \item The Rucio SKAO Datalake provides the backbone for distributed data management. It replicates SKA data products across SRC nodes according to policies of proximity, redundancy, and scientific demand, and ensures consistent namespace resolution across the federation.
    \item The SRCNet Data Management API provides programmatic access to SKAO Rucio Datalake.
    \item SRCNet IVOA DataLink acts as the interoperability layer for data and service discovery. By linking dataset replicas with registered functions and services, DataLink allows clients to query both “where the data are” and “what can be done with them,” returning the most suitable combination of site and service.
\end{itemize}

\paragraph{SRCNet Local Services}

Then, each SRC node hosts the compute and storage infrastructure required to serve its community and to integrate into the federation. Locally deployed services include interactive platforms (e.g., JupyterHub sessions, the CARTA \cite{comrie2021carta} visualisation tool, or CANFAR \cite{fabbro2024innovations} platform, among others) as well as functions such as SODA cutouts and data-preparation functions. This kind of functions are deployed as containerised services within the local SRCNet node Kubernetes cluster and connected to the local SRCNet Rucio Storage Element\footnote{Rucio RSE is logical abstraction of a storage resource, typically mapped to a physical storage system or endpoint in Rucio.} (RSE) linked with the stored dataset replicas.  

To mediate external access, a dedicated service called SRCNet GateKeeper operates at each site. It validates requests by checking the user’s IAM token and querying the SRCNet Permissions API to determine whether the requested function and dataset are authorised. Only once these checks are satisfied is the request forwarded to the appropriate service endpoint --function-- within the local Kubernetes cluster. This mechanism allows local SRC nodes to expose functions securely to the ecosystem  without compromising site security.

\begin{figure}[t]
\centering
\includegraphics[width=\textwidth]{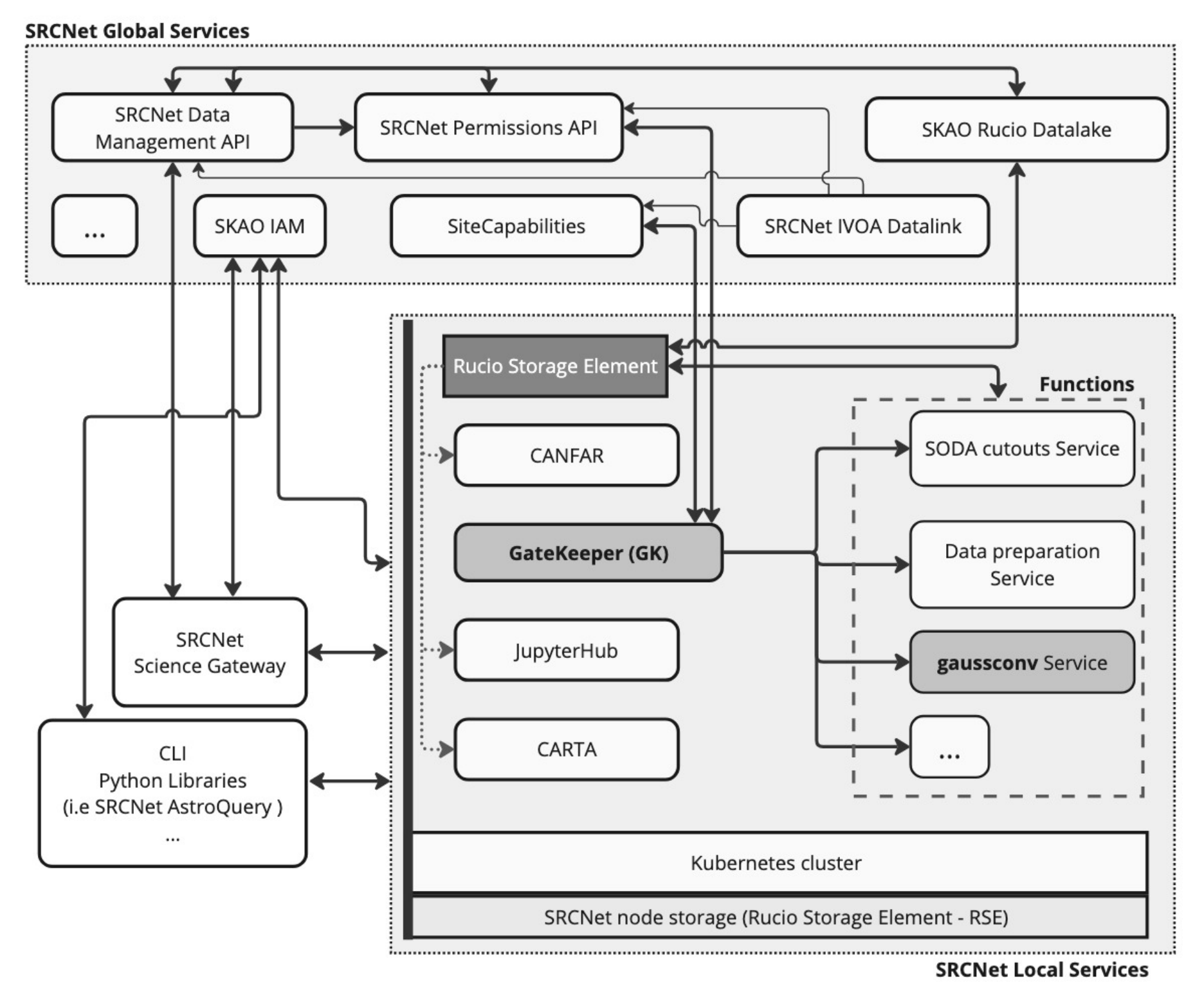}
\caption{SRCNet architecture highlighting global services (SKAO IAM, SRCNet Permissions API, SRCNet Site-Capabilities, SKAO Rucio Datalake, SRCNet IVOA DataLink) and site-local services (compute/storage infrastructure, Jupyter Notebooks/CARTA, deployed functions, \mbox{SRCNet GateKeeper}). This layered architecture provides the substrate for integrating function-oriented workflows, such as the \texttt{gaussconv} example and other.}
\label{fig:srcnet-architecture}
\end{figure}

As illustrated in Figure~\ref{fig:srcnet-architecture}, all functions deployed within a local SRC node are protected behind GateKeeper, ensuring that authentication and authorisation are enforced consistently across the federation. For example, if a user wants to run the \texttt{gaussconv} function at the Spanish SRC (espSRC) \cite{garrido2022toward}, the request must be directed to the site’s SRCNet GateKeeper endpoint rather than to the function directly. \mbox{SRCNet} GateKeeper validates the token, checks permissions, and then forwards the call to the containerised service, as shown in Listing~\ref{lst:curl-gaussconv}.

\begin{minipage}{\hsize}%
\lstset{frame=single,framexleftmargin=-1pt,framexrightmargin=-17pt,framesep=12pt,linewidth=0.98\textwidth,basicstyle=\footnotesize}
\begin{lstlisting}[caption={The user provides an OIDC access token (\texttt{\$SKA\_TOKEN}) obtained from the SKAO IAM, specifies the input dataset using its IVOA identifier, and requests a Gaussian convolution with \(\sigma=2.5\). The processed FITS file is returned and stored locally as \texttt{results.fits}. This workflow highlights the interplay between global and local services: while the function itself is deployed locally, secure and policy-compliant access is enforced by SRCNet Permissions API through the SRCNet GateKeeper mediator},label={lst:curl-gaussconv}]
curl -X POST \
 'https://gatekeeper.espsrc.iaa.csic.es/gaussconv/' \
  -H 'Authorisation: Bearer $SKA_TOKEN' \
  -H 'Content-Type: application/json' \
  -d '{
    "ivo":"ivo://espsrc.iaa.csic.es/datasets/fits?\
     testing/5b/f5/PTF10tce.fits",
    "sigma": 2.5
    }' \
 - -output /tmp/result.fits

\end{lstlisting}
\end{minipage}

Having described the overall architecture of the SRCNet and the role of its global and local services, we now focus on a concrete implementation. In the following subsection we present the \texttt{gaussconv} function as an illustrative example, detailing its development, containerisation and deployment within a local Kubernetes cluster, and its subsequent integration with the SRCNet local and global services for registration, authorisation, and discovery.

\subsection{Implementation of the \texttt{gaussconv} function within the SRCNet ecosystem} 
\label{subsec:implementation}


\begin{figure}[t]
  \centering
  \includegraphics[width=1\linewidth]{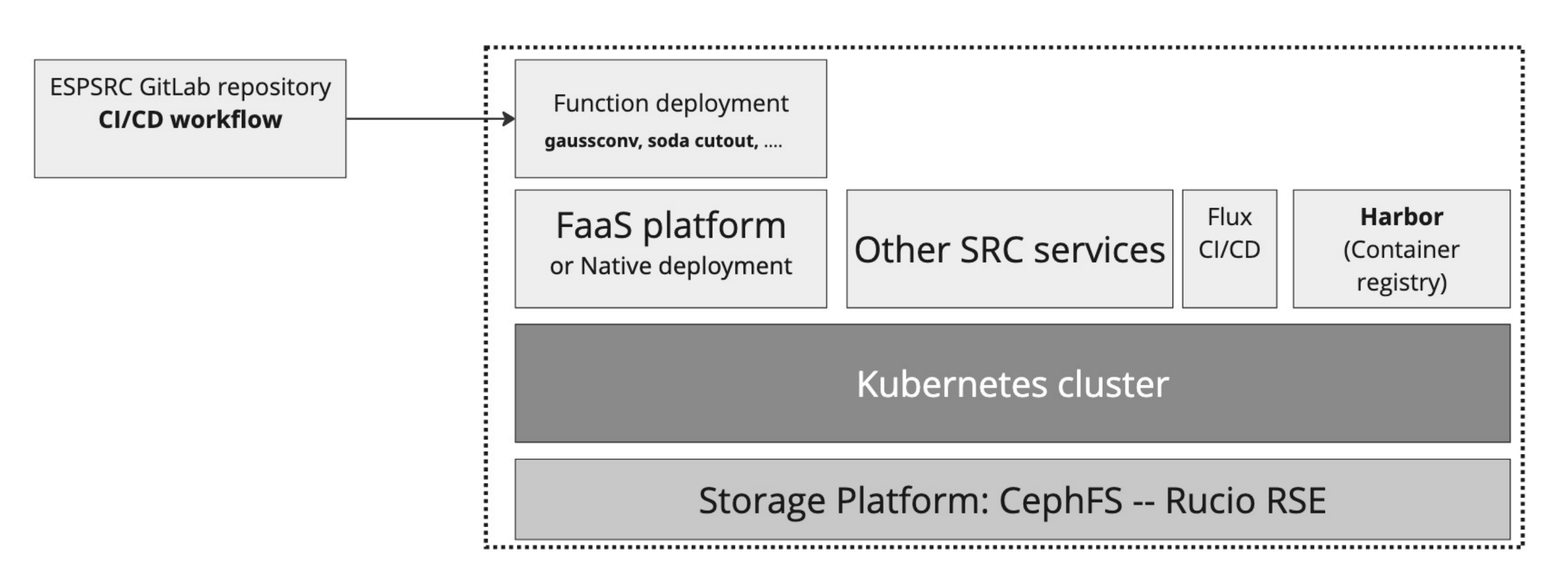}
  \caption{This figure includes the components necessary for the operation of any FaaS environment. Specifically, it focuses on the orchestration layer (Kubernetes), the storage layer for the cluster and services on the orchestrator, as well as the functions (both the FaaS platform and native mode), and the container registry, necessary for archiving container images. The automated deployment files reside in a repository in GitLab that allows CI/CD to be used for this purpose, together with a Flux CI/CD service.}
  \label{fig:arch-base}
\end{figure}

\begin{figure}[t]
\centering
\includegraphics[width=\textwidth]{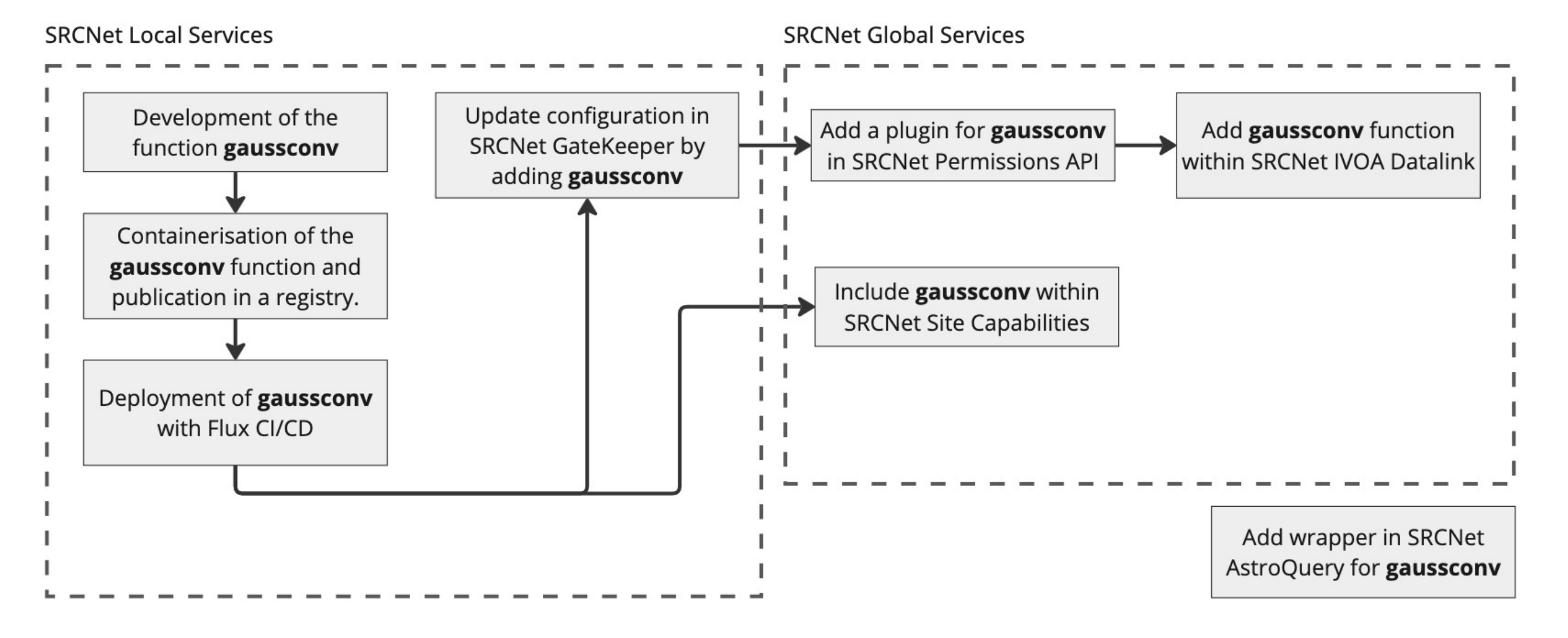}
\caption{Workflow of the steps required to integrate a new function into the SRCNet ecosystem: development and containerisation, CI/CD deployment, SRCNet GateKeeper configuration, registration in SRCNet Site-Capabilities, integration with SRCNet Permissions API and IVOA DataLink, and final exposure through the \texttt{astroquery.srcnet} wrapper.}
\label{fig:gaussconv-implementation-steps}
\end{figure}

The implementation of \texttt{gaussconv} within the SRCNet ecosystem follows a sequence of well-defined steps, summarised in Figure~\ref{fig:gaussconv-implementation-steps}. Starting from the development of the function and its containerisation, the process continues with automated deployment through CI/CD pipelines, configuration under the SRCNet GateKeeper, and registration in the SRCNet Site-Capabilities catalogue. Additional integration steps include adding a dedicated plugin in the SRCNet Permissions API to support parameter parsing, enabling discovery through SRCNet IVOA DataLink, and finally extending the \texttt{astroquery.srcnet} package with a wrapper to expose the function to end users. The SRCNet local environment required for this implementation is depicted in Figure \ref{fig:arch-base}.

\paragraph{Development of the \texttt{gaussconv} function}

At the development stage, the \texttt{gaussconv} function encapsulates a \emph{Gaussian convolution} operation over astronomical FITS images (see Listing \ref{lst:gaus_function}). The implementation is based on a Python wrapper using \texttt{FastAPI}, which provides an OpenAPI-compatible HTTP interface for parameter handling and result delivery. This design ensures that the function can be invoked through simple REST calls and easily integrated into external tools or workflows.

\begin{minipage}{\hsize}%
\lstset{frame=single,framexleftmargin=-1pt,framexrightmargin=-17pt,framesep=12pt,linewidth=0.98\textwidth,language=Python,basicstyle=\footnotesize}
\begin{lstlisting} [language=python,caption={Implementation of Gaussian convolution function.},label={lst:gaus_function}]
from fastapi import FastAPI, HTTPException, Request
from fastapi.responses import StreamingResponse
#...

app = FastAPI(title="FITS Image Convolution Service")

class GaussConvParams(BaseModel):
    """Request model for Gaussian convolution parameters."""
    ivo: str 
    sigma: float = Field(
        ...,
        ge=1.0,
        le=10.0,
        description="Sigma must be between 1 and 10"
    )
#...

async def fitsimg_gaussconv(ivo: str, sigma: float) -> bytes:
    """Logic of the function"""
    abs_path = os.getenv("ABS_PATH","/data")
    #...

@app.post("/gaussconv_fitsimg/",
  summary="Generate Gaussian convolution of a FITS image")
async def gaussconvolution_fitsimage(request: Request, 
                                params: GaussConvParams):
    #...    
    img_conv = await fitsimg_gaussconv(params.ivo, 
                                        params.sigma)
    return StreamingResponse(io.BytesIO(img_conv), 
                                media_type="image/fits")
\end{lstlisting}
\end{minipage}

A key requirement during development is the correct handling of data paths within the containerised environment. Since the function will run in a pod on a Kubernetes cluster and access dataset replicas stored locally at the SRC node (via RSE -- Rucio Storage Element), the container must be aware of the absolute path where the persistent volume is mounted. This is achieved by reading an environment variable inside the service:

\begin{minipage}{\hsize}%
\lstset{frame=single,framexleftmargin=-1pt,framexrightmargin=-17pt,framesep=12pt,linewidth=0.98\textwidth,language=Python,basicstyle=\footnotesize,}
\begin{lstlisting}[language=python,caption={Environment variable for accessing the mounted persistent volume.},label={lst:abs-path}]
import os
#...
# Absolute path to the local RSE storage
abs_path = os.getenv("ABS_PATH", "/data")
\end{lstlisting}
\end{minipage}

This mechanism allows the same container to be deployed across different SRC nodes while remaining agnostic to the specific local storage configuration. Only the environment variable needs to be set in the deployment manifest to align with the site’s mounted volume.

Once the service code is prepared, the next step is to containerise it. A dedicated \texttt{Dockerfile} specifies the base image (Python runtime), the installation of required dependencies (e.g.\ \texttt{astropy}, \texttt{numpy}, \texttt{scipy}), and the configuration of \texttt{FastAPI} together with an ASGI server (e.g.\ \texttt{uvicorn}) to expose the API. After building the image, it is tagged and pushed to a container registry, such as Harbor, from which it can be deployed on a Kubernetes service.

The full source code, service definitions, and instructions for building and publishing the image are openly available in the SRCNet GitLab repository\footnote{SKA-SRCNet \emph{Gaussian convolution} repository: \url{https://gitlab.com/ska-telescope/src/src-ia/ska-src-ia-parser-lib-gaussian-convolution}}. This repository also provides usage examples for testing the function locally, validating the container before deployment, and publishing the image to the registry.

\paragraph{Deployment of the \texttt{gaussconv} function on Kubernetes}

The deployment of the \texttt{gaussconv} function within the SRCNet testbed is managed through a GitOps workflow powered by Flux CI/CD\footnote{Flux CI/CD: \url{https://fluxcd.io/}}. Flux continuously monitors declarative configuration stored in version-controlled repositories and reconciles the live state of the Kubernetes cluster with the desired state defined in these manifests. Container image for this deployment need to be uploaded on a \emph{Harbor}\footnote{espSRC Harbor: \url{https://spsrc26.iaa.csic.es/}} service, an open-source OCI registry that enables local nodes to publish, to distribute and to manage container images. This approach ensures reproducibility and automation of the deployment process: if a new version of the container image for the function is released or any changes happened, Flux automatically detects the update and rolls out the new version to the target SRC node, managing the function’s full lifecycle.

To structure the deployment configuration, we rely on \emph{Kustomize}. \emph{Kustomize} allows declarative compositions of Kubernetes manifests, thereby enabling environment-specific customisation without duplicating code. For the \texttt{gaussconv} service, a dedicated \emph{Kustomization}\footnote{Kustomization repository for the espSRC CI/CD: \url{https://gitlab.com/ska-telescope/src/deployments/espsrc/ska-src-espsrc-services-cd/-/tree/main/apps-dev/gaussian-convolution/gitops/service?ref_type=heads}} was created that specifies the Deployment, Service, Storage and Volumes required for the function to operate.

Once the deployment is running within the Kubernetes cluster, the \texttt{gaussconv} function is available only as an internal service (\texttt{ClusterIP}) within a gausconv namespace. At this stage, it can be invoked locally inside the cluster but is not yet accessible from external clients. To make the function consumable by the wider SRCNet ecosystem---for example from APIs, command-line tools, or Python packages such as \texttt{astroquery.srcnet}---additional integration steps are required.  

\paragraph{SRNet GateKeeper and function registration in Site Capabilities}

First, the function must be proxied through the SRCNet GateKeeper, which acts as the single entry point for all external function invocations as describe in Figure \ref{fig:srcnet-architecture}. GateKeeper enforces authentication and authorisation policies by validating IAM tokens and validating the SRCNet Permissions API, before securely forwarding the request to the local service endpoint. This guarantees that only authorised users can invoke the function on the datasets to which they have access. SRCNet GateKeeper configuration is declarative: each function is defined by its route, namespace, service name, exposed port, and unique identifier. For the \texttt{gaussconv} function, the corresponding entry in the GateKeeper configuration is as follows:

\begin{minipage}{\hsize}%
\lstset{frame=single,framexleftmargin=-1pt,framexrightmargin=-17pt,framesep=12pt,linewidth=0.98\textwidth,language=xml,basicstyle=\footnotesize}
\begin{lstlisting}[language=bash,caption={This configuration ensures that any external call to \texttt{/gaussconv} is redirected by SRCNet GateKeeper to the \texttt{gaussconv-srv} service running in the \texttt{gaussconv} namespace on port~8000, provided that the user is authenticated and authorised.},label={lst:gk-config}]
service:
  - route: "/gaussconv"
    namespace: "gaussconv"
    service_name: "gaussconv-srv"
    ingress_host: ""
    port: 8000
    uuid: "<redacted>"
\end{lstlisting}
\end{minipage}

\begin{figure}[t]
\centering
\includegraphics[width=0.9\textwidth]{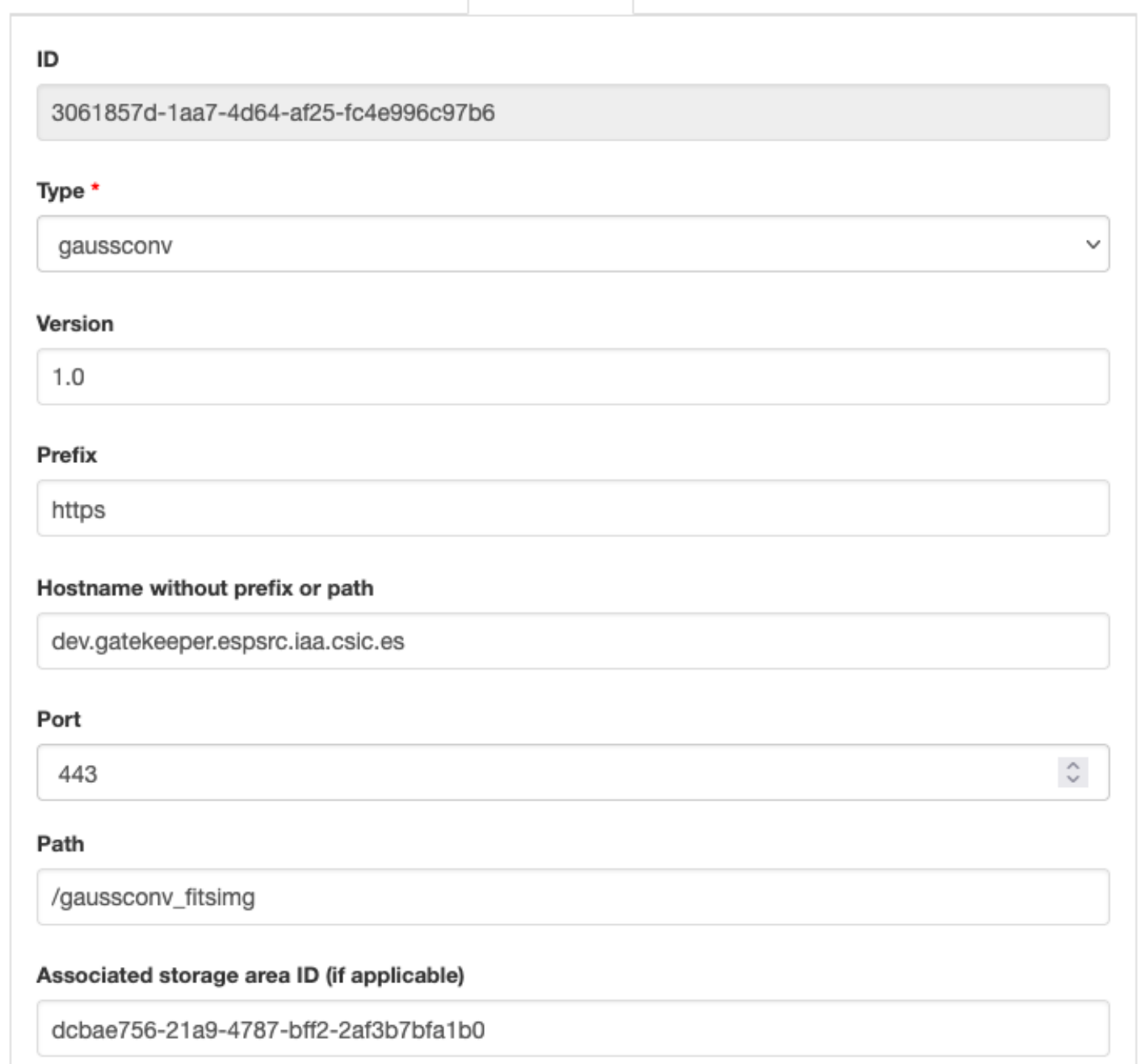}
\caption{Example of registering the \texttt{gaussconv} function in the SRCNet Site-Capabilities catalogue. Metadata such as the function UUID, prefix, supported parameters, and endpoint are added to enable federation-wide discovery and invocation.}
\label{fig:site-capabilities}
\end{figure}

Second, the function must be registered in the \mbox{SRCNet} Site-Capabilities catalogue\footnote{SRCNet Site-Capabilities API and documentation: \url{https://site-capabilities.srcnet.skao.int/api/v1/www/docs/oper}}. This registration process makes the function discoverable across the federation, associating it with metadata such as its unique identifier (UUID), the RSE storage associated, the port and the endpoint where it can be reached (through \mbox{SRCNet} GateKeeper), among others. Registration in SRCNet Site-Capabilities ensures that higher-level services, such as IVOA DataLink, can advertise the function alongside the datasets it operates on.   Figure~\ref{fig:site-capabilities} illustrates the process of adding a new record to the Site-Capabilities service for the \texttt{gaussconv} function, thereby making it discoverable and callable across the SRCNet federation.

\paragraph{Integration with the Permissions API and SRCNet IVOA Datalink}

For SRCNet GateKeeper to validate whether a user is authorised to access a specific dataset and namespace in the SKAO Rucio Datalake, it is necessary to extend the SRCNet Permissions API with a dedicated plugin. This plugin encapsulates the logic for extracting the namespace from the dataset identifier provided in the request, thereby allowing SRCNet GateKeeper to query the SRCNet Permissions API and confirm access rights.  
The plug-in must be added to the permissions API codebase under the \texttt{plugins} directory\footnote{SRCNet permissions API plugins folder: \url{https://gitlab.com/ska-telescope/src/src-service-apis/ska-src-permissions-api/-/tree/main/src/ska_src_permissions_api/plugins?ref_type=heads}}, enabling tailored policies for functions such as \texttt{gaussconv}. With this mechanism, the authorisation process ensures that both the dataset and the requested operation are validated consistently across the federation.

The IVOA DataLink standard provides the mechanism to associate datasets with related services or processing functions. Within SRCNet, this allows users not only to discover where dataset replicas are stored, but also which functions are available to operate on them. To support the \texttt{gaussconv} function, the DataLink service template must be extended to include the function metadata and its input parameters in the returned VOTable. A simplified template fragment is shown in Listing \ref{lst:datalink-gaussconv}:

\begin{minipage}{\hsize}%
\lstset{frame=single,framexleftmargin=-1pt,framexrightmargin=-17pt,framesep=12pt,linewidth=0.98\textwidth,basicstyle=\footnotesize}
\begin{lstlisting}[language=xml,caption={VOTable snippet for exposing \texttt{gaussconv} in IVOA DataLink.},label={lst:datalink-gaussconv}]
{% if include_services_flags.gaussconv %}
<RESOURCE type="meta" ID="gaussconv" utype="adhoc:service">
  <PARAM name="resourceIdentifier" datatype="char" 
    arraysize="{{ services.gaussconv.resource_identifier | length }}" 
    value="{{ services.gaussconv.resource_identifier }}" />
  <PARAM name="accessURL" datatype="char" 
    arraysize="{{ services.gaussconv.access_url | length }}" 
    value="{{ services.gaussconv.access_url }}" />
  <GROUP name="inputParams">
    <PARAM name="ID" datatype="char" 
      arraysize="{{ ('ivo://' ~ ivoa_authority ~ '?' ~ path_on_storage) |
      length }}" 
      ucd="meta.id;meta.dataset" 
    value="ivo://{{ ivoa_authority }}?{{ path_on_storage }}" />
    <PARAM name="SIGMA" datatype="double" value="1.0" />
  </GROUP>
</RESOURCE>
{% endif %}
\end{lstlisting}
\end{minipage}

\paragraph{Integration with \texttt{astroquery.srcnet}}

Finally, the \texttt{astroquery.srcnet} Python client encapsulates the full workflow for users, hiding the complexity of the SRCNet ecosystem. To achieve this, a dedicated wrapper for the \texttt{gaussconv} function is required, which constructs the DataLink query with the appropriate parameters, and executes the function at the selected SRC node. An example of its usage is shown below:

\begin{minipage}{\hsize}%
\lstset{frame=single,framexleftmargin=-1pt,framexrightmargin=-17pt,framesep=12pt,linewidth=0.98\textwidth,language=xml,basicstyle=\footnotesize}
\begin{lstlisting}[language=python,caption={Example usage of the \texttt{gaussconv} wrapper in \texttt{astroquery.srcnet}.},label={lst:astroquery-gaussconv}]
from astroquery.srcnet import SRCNet

# Authenticate with SKAO IAM
SRCNet.login()

# Invoke Gaussian convolution on a FITS dataset
result = SRCNet.gaussconv(
    namespace="testing",
    name="PTF10tce.fits",
    output_file="output/PTF10tce_gaussconv_sigma2.5.fits",
    sigma=2.5,
)
\end{lstlisting}
\end{minipage}

\section{Discussion}
\label{sec:discussion}

\subsection{Comparison with existing FaaS platforms}

The model proposed for the SRCNet preserves the essence of Function-as-a-Service but differs substantially from the platforms commonly used in commercial or single-cloud contexts. Services such as AWS Lambda, OpenFaaS, or Knative provide effective mechanisms for elastic execution of stateless functions, but they assume a centralised infrastructure with homogeneous control of identity, resources, and data. In these systems, functions are deployed within a single cloud provider or cluster, registered in a local catalogue, and authorised using the provider’s internal IAM policies. While this model simplifies development for enterprise applications, it does not address the requirements of federated scientific infrastructures.

In contrast, the SRCNet model is designed for a heterogeneous network of regional centres, each with its own storage, compute resources, and access policies, but all federated under a common identity and discovery framework. SRCNet Site-Capabilities extends the notion of a function catalogue by linking function deployments to specific data namespaces and by embedding policy references for authorisation. SRCNet GateKeeper introduces an explicit enforcement layer that validates the user identity and permissions before routing any request to the internal cluster services. Finally, SRCNet IVOA DataLink provides standardised discovery of both data replicas and available functions, enabling data-proximate execution across geographically distributed sites. These mechanisms are not add-ons but fundamental requirements in a scenario where exascale datasets are distributed across multiple institutions and need to be processed close to where they are stored.

\begin{table}[t]
  \centering
  \caption{Comparison between existing FaaS platforms and the SRCNet function-oriented model.}
  \label{tab:faas-comparison}
  \begin{tabular}{p{0.25\linewidth}p{0.3\linewidth}p{0.3\linewidth}}
    \toprule
    Feature & Generic FaaS (AWS Lambda, OpenFaaS, Knative) & SRCNet Function-Oriented Model \\
    \midrule
    Execution scope & Single cloud or cluster & Federated across SRC nodes \\
    Identity & Provider-specific IAM & OIDC via SKAO IAM \\
    Function registry & Local catalogue per cluster & Site-Capabilities (federated, policy-aware) \\
    Data access & Data copied to function & Bound to namespaces, executed close to data \\
    Discovery & Limited or ad-hoc metadata & IVOA DataLink with function/data coupling \\
    Security & Provider IAM policies & GateKeeper with user–function–namespace checks \\
    Target domain & Enterprise and web apps & Data-intensive astronomy (SKA, SRCNet) \\
    \bottomrule
  \end{tabular}
\end{table}

\subsection{Impact on scientific data processing}

The most direct benefit of this model lies in its capacity to reduce data movement, which is one of the central bottlenecks in large astronomical projects. By ensuring that functions are discovered and executed at the site where the corresponding dataset resides, the system avoids costly transfers across the federation and minimises duplication of intermediate products. This principle of data-proximate execution is crucial in the SKA and SRCNet context, where raw and processed data products will be distributed across multiple regional centres and cannot be feasibly centralised.

This approach also paves the way for the construction of federated libraries of functions that can be developed once and applied across all SRC nodes. Researchers gain access to a consistent catalogue of operations—such as filtering, calibration, or data transformation—that are guaranteed to be executed within the policy and security framework of the SRCNet. This increases reproducibility of analyses, since the same function with the same version can be invoked independently of the site where the data is located. Moreover, by embedding metadata about supported data types and namespaces, the model ensures that only compatible functions are presented to users, reducing errors and promoting standardisation in scientific workflows.

These features align with broader goals in scientific computing, such as adherence to the FAIR principles. Functions become not only reusable but also findable through DataLink, accessible under federated identity, interoperable across heterogeneous resources, and reproducible through versioned registration in SRCNet Site-Capabilities. In addition, the function code is available in public repositories.

\subsection{Limitations of the current model}

Despite its advantages, the proposed model introduces several limitations that must be acknowledged. The first is architectural complexity. In contrast to commercial FaaS environments, where deployment and invocation are largely handled by a single provider, the SRCNet model requires coordination across multiple services: SRCNet Site-Capabilities for registration, SKAO IAM and SRCNet Permissions for policy enforcement, SRCNet GateKeeper for secure mediation, and SRCNet  DataLink for discovery. This additional overhead is necessary in a federated scientific setting, but increases the operational management for site administrators and the cognitive load for developers. Ensuring that all services remain consistent and available across heterogeneous SRC nodes is itself a challenge.

A second limitation lies in the orchestration of workflows composed of multiple functions. The current design is optimised for single function invocations, where inputs and outputs are tied directly to persistent datasets. However, many scientific analyses require chaining several operations, with intermediate results that may be transient and not suitable for long-term storage. At present, such workflows require materialising intermediate products, incurring latency and storage costs. Efficient support for ephemeral data exchange between functions, without compromising security and provenance, remains an open problem\footnote{For example, chaining a \emph{Gaussian convolution} step with subsequent noise removal requires persisting an intermediate FITS cube, which could otherwise be streamed directly between functions.}.

Finally, heterogeneity across SRC nodes poses additional limitations. Not all sites may support the same resource types or capabilities, leading to variability in performance and availability. While the architecture allows functions to be deployed at multiple sites, maintaining consistent versions, policies, and parameter schemas across the federation requires careful synchronisation. These issues highlight that the model, while functional, is still evolving and it will need further refinement before adoption at SRCNet scale.

\subsection{Broader implications and generalisation}

The design principles developed here are not limited to astronomy or to the \mbox{SRCNet}. The combination of federated identity, policy-aware function registration, and data-proximate execution addresses challenges common to many large-scale scientific projects. Infrastructures such as the LHC, the Cherenkov Telescope Array, or the upcoming Euclid mission face similar requirements: exascale datasets distributed across international centres, data governance rules, and the need for reproducible, federated computation. The SRCNet model demonstrates how FaaS concepts can be adapted to these environments, potentially serving as a blueprint for cross-domain adoption.

\begin{figure}[h]
\centering
\includegraphics[width=1\textwidth]{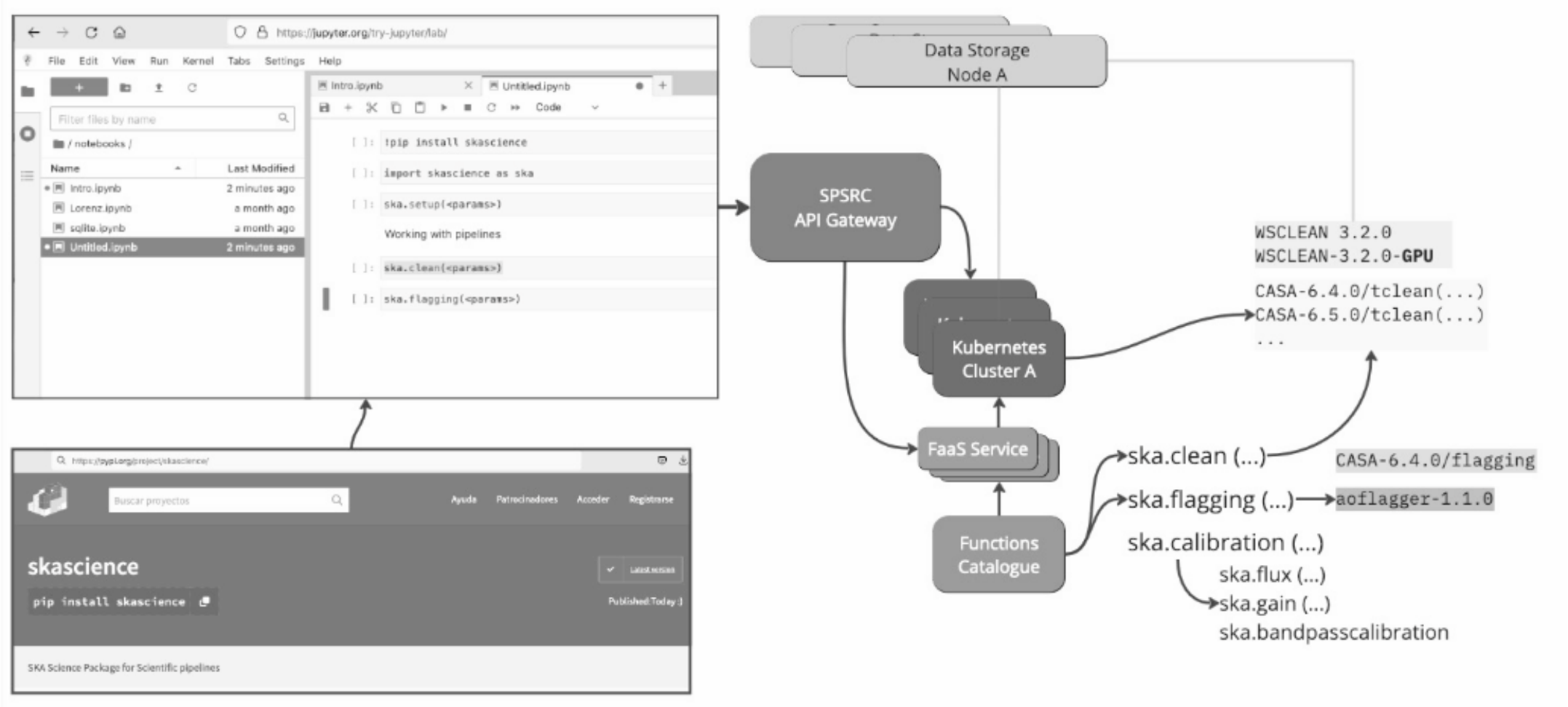}
\caption{Generic environment for a radio-astronomy serverless catalogue model, publication of functions in a Kubernetes cluster, connection to data storage, and Python package so that scientists can use the catalogue of functions from Jupyter Notebooks or within a script.}\label{fig:serverless-gateway}
\end{figure}

Beyond scientific projects, the approach also opens opportunities for more general federated function marketplaces. By standardising the publication of functions with metadata about data types, supported architectures, and security requirements, communities could share curated libraries of operations across institutional boundaries. Discovery services such as DataLink could evolve to select not only the closest data replica, but also functions optimised for specific hardware such as GPUs, ARM, FPGAs, or other accelerators. Figure \ref{fig:serverless-gateway} presents several images for the same function for different computing architectures and versions. This would enable resource-aware scheduling where astronomers or scientists simply declare the desired operation, while the infrastructure automatically routes the request to the most suitable site and function.

The implications extend further to policy-driven execution. In addition to proximity or hardware affinity, discovery mechanisms could consider cost models, jurisdictional restrictions, or project-level priorities when selecting a function endpoint. By embedding these dimensions into federated catalogues and discovery protocols, the SRCNet approach lays the groundwork for a new generation of distributed computing environments.

\section{Conclusions and Future Work}
\label{sec:conclusion}

This work has demonstrated the feasibility of adapting FaaS to the context of the SRCNet. The proposed function-oriented architecture preserves the lightweight, containerised nature of FaaS while extending it with the mechanisms required for a federated scientific infrastructure. By integrating the function into the SRCNet ecosystem, the model provides the necessary foundation for secure, traceable, and data-proximate execution of functions across all SRC nodes.

The implementation of the \texttt{gaussconv} function illustrates how scientific operations can be containerised, deployed on local Kubernetes clusters, and exposed through SRCNet GateKeeper for federation-wide access. Coupling functions directly with dataset replicas reduces network transfers and latency, enabling workflows to operate efficiently on large, geographically distributed data volumes. This design also opens the door to curated libraries of reusable functions, promoting reproducibility and consistency, and to exploiting hardware diversity by selecting functions based on proximity as well as resource descriptors such as GPU acceleration among others.

Several challenges remain. The current model focuses on single-function execution; orchestrating chained pipelines where outputs of one function serve as inputs for others remains an open issue, particularly for handling transient intermediate data securely and efficiently. Additional work is also needed on fault tolerance, adaptive scheduling across sites, and provenance tracking to ensure robustness at SKA scale.

Future work will extend the architecture in three directions. First, enabling function chaining and workflow composition to support complex pipelines without unnecessary data materialisation. Second, enriching scheduling and discovery so that SRCNet DataLink can account for proximity, resource availability, and hardware accelerators, for example. Third, strengthening reproducibility through provenance metadata, versioning, and policy-driven management of trasient data. Together, these steps will consolidate FaaS as a practical foundation for federated, function-oriented computation in the SRCNet and, ultimately, for the data-intensive workflows of the SKA context.

\section*{Statements and Declarations}

\subsection*{Competing Interests}
The authors declare that they have no competing interests.

\subsection*{Funding Information and acknowledgements}
MP, SSE, JG, JS, MM, LD, EJ and LVM acknowledge financial support from the grant CEX2021-001131-S funded by MICIU/AEI/ 10.13039/501100011033 and from the grant TED2021-130231B-I00 funded by MICIU/AEI/  and by the European Union NextGenerationEU/PRTR, acknowledges financial support from the grants PID2021-123930OB-C21 and PID2024-155817OB-I00 funded by MICIU/AEI/  and by ERDF/EU, and by the grant INFRA24023 (CSIC4SKA) funded by CSIC. Prototype of an SRC (SPSRC) service and support funded by the Ministerio de Ciencia, Innovación y Universidades (MICIU), by the Junta de Andalucía, by the European Regional Development Fund (ERDF) and by the European Union NextGenerationEU/PRTR. The SPSRC acknowledges financial support from the Agencia Estatal de Investigación (AEI) through the "Center of Excellence Severo Ochoa" award to the Instituto de Astrofísica de Andalucía (IAA-CSIC) (SEV-2017-0709) and from the grant CEX2021-001131-S funded by MICIU/AEI. AM acknowledges support from the UK SKA Regional Centre (UKSRC). The UKSRC is a collaboration between the University of Cambridge, University of Edinburgh, Durham University, University of Hertfordshire, University of Manchester, University College London, and the UKRI Science and Technology Facilities Council (STFC) Scientific Computing at RAL. The UKSRC is supported by funding from the UKRI STFC.

\subsection*{Author Contribution}
M.~Parra (MP), S.~Sánchez-Expósito (SSE), Julián Garrido (JG) and Jesús Sánchez conceived this research.  MP wrote the first version of the manuscript in its entirety; all figures, tables and diagrams were also produced by MP. The article underwent extensive review by SSE and JG, with additional revisions by L.~Darriba (LD) and M.~Ángeles. M.~Ángeles also contributed with the function code and by updating the relevant repositories. Anthony  Moraghan developed the astroquery package for the SRCNet, MP developed the SRCNet permissions API and Rob Barnsley the SRCnet IVOA datalink service. The remaining co-authors contributed by reviewing the document and/or by developing services and tools that form the SRCNet ecosystem used in this study. JG, SSE and Lourdes Verdes-Montenegro funded this research. 

\subsection*{Data and Code Availability}
All code used in this work is available in public SRCNet repositories. The source for the function implementations can be found at: \url{https://gitlab.com/ska-telescope/src/src-ia/ska-src-ia-parser-lib-gaussian-convolution}.
Repository links for the wider ecosystem referenced herein are:
\begin{itemize}
  \item SRCNet IVOA DataLink: \url{https://gitlab.com/ska-telescope/src/src-mm/ska-src-mm-rucio-ivoa-integrations} 
  \item SRCNet Permissions API: \url{https://gitlab.com/ska-telescope/src/src-service-apis/ska-src-permissions-api/-/tree/main/src/ska_src_permissions_api?ref_type=heads} 
  \item \texttt{astroquery.srcnet} client: \url{https://gitlab.com/ska-telescope/src/src-ui/ska-src-astroquery}
\end{itemize}
Where applicable, example configuration files and deployment manifests (Kubernetes/CI--CD) are included in the repositories. No proprietary datasets were used; function executions operate within SRCNet infrastructure.

\subsection*{Research Involving Humans and/or Animals}
Not applicable.

\subsection*{Informed Consent}
Not applicable.

\bibliography{sn-bibliography}

\end{document}